%% file: main.tex
\documentclass[10pt,conference]{IEEEtran}
\usepackage{geometry}
\PassOptionsToPackage{table,xcdraw}{xcolor}
\usepackage[utf8]{inputenc}
\usepackage[T1]{fontenc}

\usepackage{hyperref}

\newcommand{\ra}[1]{}

\usepackage{graphics} 
\usepackage{epsfig} 
\usepackage{amsmath} 
\usepackage{bbm}
\usepackage{dsfont}
\usepackage{multirow}
\usepackage{subcaption}
\usepackage{pgfplots}

\usepackage{amsbsy}
\usepackage{amsfonts}
\usepackage{array, booktabs, longtable}
\usepackage{graphicx}
\usepackage{fixltx2e}
\usepackage{color}
\usepackage{algpseudocode, algorithm}
\makeatletter
\renewcommand{\ALG@beginalgorithmic}{\small}
\makeatother

\begin{document}



%


\newcommand\blfootnote[1]{%
  \let\thefootnote\relax\footnotetext{#1}%
  \let\thefootnote\svthefootnote%
}
\renewcommand{\footnoterule}{%
  \kern -3pt
  \hrule width 2cm
  \kern 2pt
}

\title{\Large\bf Net Separation-Oriented Printed Circuit Board Placement via Margin Maximization}

	\author{\normalsize
		\large Chung-Kuan Cheng \large Chia-Tung Ho  and Chester Holtz\IEEEauthorrefmark{1} \\ \\
		\{ckcheng, c2ho, chholtz\}@eng.ucsd.edu
		\\ Department of Computer Science and Engineering \\
		University of California San Diego\\ La Jolla, CA, 92037 \\
}

\maketitle

\makeatletter
\def\ps@IEEEtitlepagestyle{%
  \def\@oddfoot{\mycopyrightnotice}%
  \def\@evenfoot{}%
}
\makeatother
\def\mycopyrightnotice{%
  \begin{minipage}{\textwidth}
    \footnotesize
    ~ \hfill\\~\\
  \end{minipage}
  \gdef\mycopyrightnotice{}
}

{\small\bf Abstract---\blfootnote{*Corresponding author}
Packaging has become a crucial process due to the paradigm shift of More than Moore. Addressing manufacturing and yield issues is a significant challenge for modern layout algorithms. \\
\indent We propose to use printed circuit board (PCB) placement as a benchmark for the packaging problem. A maximum-margin formulation is devised to improve the separation between nets. Our framework includes seed layout proposals, a coordinate descent-based procedure to optimize routability, and a mixed-integer linear programming method to legalize the layout. We perform an extensive study with 14 PCB designs and an open-source router. We show that the placements produced by NS-place improve routed wirelength by up to 25\%, reduce the number of vias by up to 50\%, and reduce the number of DRVs by 79\% compared to manual and wirelength-minimal placements.}

\input{Introduction.tex}
\input{RelatedWork.tex}
\input{Formulation.tex}

\input{Experiments.tex}
\input{Conclusion.tex}
\bibliographystyle{IEEEtran}
{\footnotesize
\bibliography{main}
}

\end{document}

%% file: Introduction.tex
\section{Introduction}
\label{sec:introduction}

Packaging technology continues to advance from Printed Circuit Boards, Flip-Chip, Integrated Fan-Out (InFO), Chip-on-Wafer-on-Substrate (CoWoS)~\cite{douglas2014wafer}, Integrated Fan-Out (InFO)~\cite{douglas2017advanced}, 
to System-on-Integrated-Circuit (SoIC)~\cite{chen2019system} for More than Moore. However, the integration of the packages encounters significant manufacturability and yield issues due to components with arbitrary shape, non-Manhattan routing directions, vias of larger than routing track pitches, high resistance, and/or reliability problems. In this paper, we use Printed Circuit Board placement as a test vehicle to improve manufacturability and yield. We separate nets in 2D space to minimize net crossings, and encourage same layer routing. Our goal is to minimize post-route design rule violations, and reduce via usage. We additionally expect a reduction in the routed metal length.

The PCB placement problem exhibits additional challenges compared to Integrated Chip (IC) placement including arbitrary component shapes,  board boundaries, multiple layers, and routing constraints. Conventional analytical IC placers \cite{ckcheng2018replace} suffer from several key issues which prevent convergence to good\textemdash or even reasonable\textemdash solutions (with respect to wirelength and routability): (1) The vanilla RePlace \cite{ckcheng2018replace} implementation fails to find good solutions for multi-layer designs with diversely sized components. (2) The rate of convergence of the algorithm is dependent on design-specific parameters\textemdash e.g. filler cells and anchor weights.

In this work, we propose and apply a gradient descent-based algorithm to optimize wirelength, component density, and routability and a mixed-integer linear program (MILP) for local legalization. The computational bottleneck introduced by the MILP is addressed by integrating global relative positioning constraints derived from the gradient-based placement stage. Furthermore, our framework involves very few design-dependent parameters. This allows our framework to be generally applied to a variety of designs without a costly tuning stage. 
\subsection{Contribution}
Our contributions can be summarized as follows:
\begin{enumerate}
\itemsep=0em
    \item We propose the NS-Place framework for PCB layout which minimizes net congestion using a support vector machine-like formulation and performs legalization by solving a congestion-aware MILP.
    \item We demonstrate that the routed placements produced by our framework have fewer design rule violations and vias, and shorter total metal length compared to manual placements.
\end{enumerate}
In section~\ref{sec:related}, we review previous work. In section~\ref{sec:formulation}, we describe our routability objective. The NS-Place placement framework, including initialization and the MILP-based legalizer is described in Section~\ref{sec::flow}. In section~\ref{sec:results}, we present experimental results on real PCB testcases. We conclude in section~\ref{sec:conclusion}.

%% file: RelatedWork.tex
\section{Related Work}
\label{sec:related}

In this section we review the previous work with respect to cost and congestion-driven placement and PCB layout.

\subsection{Cost-driven placement}
Conventional global placement strategies for ICs seek to minimize wirelength subject to density constraints. Density constraints are typically integrated with the objective to yield an unconstrained relaxation (e.g. as in \cite{ckcheng2018replace}):
\begin{equation}
  \begin{aligned}
    \min_{x,y} (\sum_{e=(i,j) \in \mathcal{E}} Wl(e; x, y) + \lambda D(x,y))
  \end{aligned}
  \label{eq::repl}
\end{equation}
where $Wl(\cdot; \cdot)$ is a function that takes a net instance $e$ as input and returns the cumulative wirelength,, and $D(\cdot)$ is a density penalty. In the context of IC layout, the wirelength of a net is commonly modelled as the half-perimeter wirelength (HPWL) or a smooth alternative and $D$ is a smooth density \cite{LuePlace15}. Overlap constraints are typically satisfied over the placement process by gradually increasing the weight $\lambda$, at the cost of increased wirelength. The current state-of-the-art IC placement algorithms \cite{LuePlace15, ckcheng2018replace} solve Problem \ref{eq::repl} in this manner. We adopt a similar formulation for PCB placement.
\subsection{Congestion-driven Placement}

A typical method of estimating routing demand is to consider pin or feasible routed-wire density~\cite{spindler2007rudy}. Other methods include applying Rent's rule~\cite{li04}, or more sophisticated routing models; for example relying on the construction of rectilinear Steiner trees or the external evaluation of a router \cite{roy07}. State of the art techniques for congestion-aware placement include mPL~\cite{li04}, a multilevel analytical placer based on non-linear optimization and estimating the routing demand based on a two-pin connection routing model, ROOSTER~\cite{roy07}: a min-cut placer which models nets by Rectilinear Steiner Minimal Trees, and APlace~\cite{kahngaplace}, a multilevel analytical placer based on non-linear optimization and stochastic estimates of the routing demand. Similar to our work, \cite{shabbeer2012crossing} propose to minimize a smooth upper bound on the crossing number to reduce edge crossings in the context of graph visualization, but their formulation is incapable of handling multi-pin nets.

These techniques generally suffer from inadequate estimation or prohibitive computational cost. In contrast, the framework proposed in this work is rigorous and does not rely on rerunning the placement algorithm or applying post-placement optimization. 

\subsection{PCB Placement}
Examples of previous work on the PCB placement problem include \cite{JainPCBGenetic96, FatimahSOGAA12, BadriyahGARLEE16, AlexSwarm17}. These techniques rely on various meta-heuristics to produce non-overlapping layouts while taking into account various metrics such as thermal and power characteristics of components, timing, and tidiness. In general, these methods suffer from drawbacks\textemdash e.g. are computationally expensive, incapable of rotating components, or evaluated on synthetic or toy benchmarks. In contrast, our framework is efficient, capable of rotating modules, extensible, and validated on production PCB designs placed by industry experts. We additionally acknowledge the similarity of the PCB placement problem to macro placement, and point the reader to \cite{AdyaMacroPlace05} for a review of relevant techniques.

%% file: Formulation.tex
\section{Net Separation-Oriented Placement}
\label{sec:formulation}

\subsection{Preliminaries}
%
\begin{figure}[!h]
\fbox{\begin{minipage}[t]{0.45\textwidth}
    net \hfill $e\in\mathcal{E}$ \\
    pin matrices \hfill  $A_e \in \mathbb{R}^{p\times2}$ \\
    coordinates and orientation \hfill $x,y \in \mathbb{R}_+^n$, $r \in \{0,1\}^n$ \\
    density \& net separation cost weight \hfill $\lambda_{\textrm{D}}, \lambda_{\textrm{NS}} \in 
    \mathbb{R}_+$ \\
    convex hull coefficients \hfill $u\in \mathbb{R}^2, \gamma\in \mathbb{R}$ \\
    wirelength smoothing parameter \hfill $c \in \mathbb{R}_+$
\end{minipage}}
\caption{Notation \& key terms}
\label{tab:hparams}
\end{figure}
%
Let $x, y\in \mathbb{R}_+^{n}$ be vectors corresponding of coordinates of $n$ components such that the $i$-th component has coordinates encoded in the $i$-th row of $[x:y]$; $[x:y]_i$. Let $\mathcal{E}$ denote a set of $m$ nets. We aim to assign coordinates so that the resulting layout has small cumulative wirelength, layout density, and routing congestion. 
\subsection{NS-Place Objective}
Our method may be expressed concisely as the following unconstrained optimization problem given $\lambda$:
\begin{equation}
    \min_{x, y} \sum_{e\in\mathcal{E}}\left[Wa(e;x,y) + \lambda_{\textrm{NS}}Ns(e;x,y)\right] +\lambda_{\textrm{D}} D(x,y)
    \label{eq:obj}
\end{equation}
where $Wa$ and $D$ corresponds to weighted-average wirelength and density terms, and $Ns$ corresponds to the proposed net separation term described in Sec.~\ref{sec:formulation}.D.

\subsection{Wirelength and density-driven optimization}

Many modern techniques for analytic placement rely on quadratic optimization with terms associated with attraction of connected cells, and repulsion of overlapping cells. A typical approach is to represent individual nets as rectangles and to minimize the sum-perimeters over all nets. Repulsion is often applied between overlapping nodes to reduce density. In this work, we adopt the smooth continuous and differentiable weighted-average wirelength (Wa) model~\cite{hsuHPWL} for wirelength cost. The horizontal net-wirelength for net $e$ is given by
$$
Wa^{(e)}_x = \frac{\sum_{i\in e}x_i\exp{(\frac{x_i}{c})}}{\sum_{i\in e}\exp{(\frac{x_i}{c})}} - \frac{\sum_{i\in e}x_i\exp{(-\frac{x_i}{c})}}{\sum_{i\in e}\exp{(-\frac{x_i}{c})}}
$$
where $c$ is a parameter that controls the smoothness and approximation error.
%
%
%
We then write the wirelength of $e$: 
$$
Wa(e;x,y) = Wa^{(e)}_x + Wa^{(e)}_y
$$
The density term corresponds to the mixed-size module bin-based density objective described in \cite{kahngaplace}. The placement area is divided into $B$ bins, and the placer seeks to equalize the overlap at each bin. For a bin $b$, let $x_b$ be the $x$-coordinate of the center and $w_b$ be the width. Then the smoothed overlap $\Theta_x(b, i)$ in the $x$-direction between bin $b$ and module $i$ with width $w_i$ and height $h_i$ is
$$    
\Theta_x (b,i) = 
    \begin{cases}
        1-2d_x^2/w_b^2,\quad \textrm{if }0 \leq d_x \leq w_b/2\\
        2(d_x - w_b)^2/w_b^2,\quad \textrm{if }w_b/2 \leq d_x \leq w_b\\
        0 \quad \textrm{if }w_b \leq d_x
    \end{cases}
$$
where $d_x = |x_i - x_b|$. The overlap in the $y$-direction is defined similarly. The density function of bin $b$ is then
\begin{equation}
    D_b(x,y) = \sum_{i}C_i \Theta_x(b,i)\Theta_y(b,i)
\end{equation}
where $C_i$ is a normalization factor such that $\sum_{b}C_i\Theta_x(b,i)\Theta_y(b,i) = w_i h_i$\textemdash the area of module $i$. Finally, $D(x,y) = \sum_{b}(D_b(x,y) - \frac{\sum_i (w_ih_i)}{B})^2$.

\subsection{Net-separation optimization via margin maximization}
\begin{figure}[t]
\centering
\begin{minipage}[b]{.48\linewidth}
    \centering
    \includegraphics[width=0.8\linewidth,trim=4 4 4 4,clip]{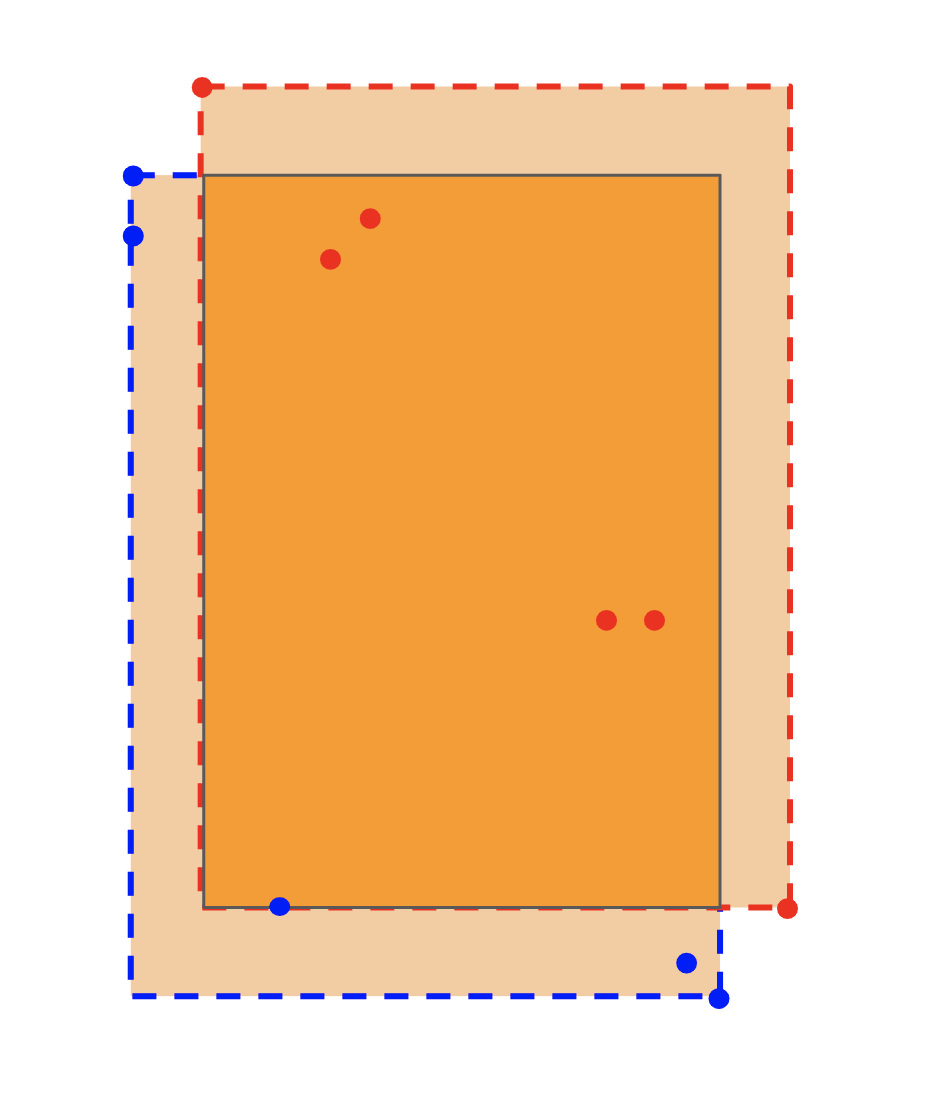}
    \subcaption{}
    \label{fig:rudy}
\end{minipage}%
\begin{minipage}[b]{.48\linewidth}
    \centering
    \includegraphics[width=0.8\linewidth,trim=4 4 4 4,clip]{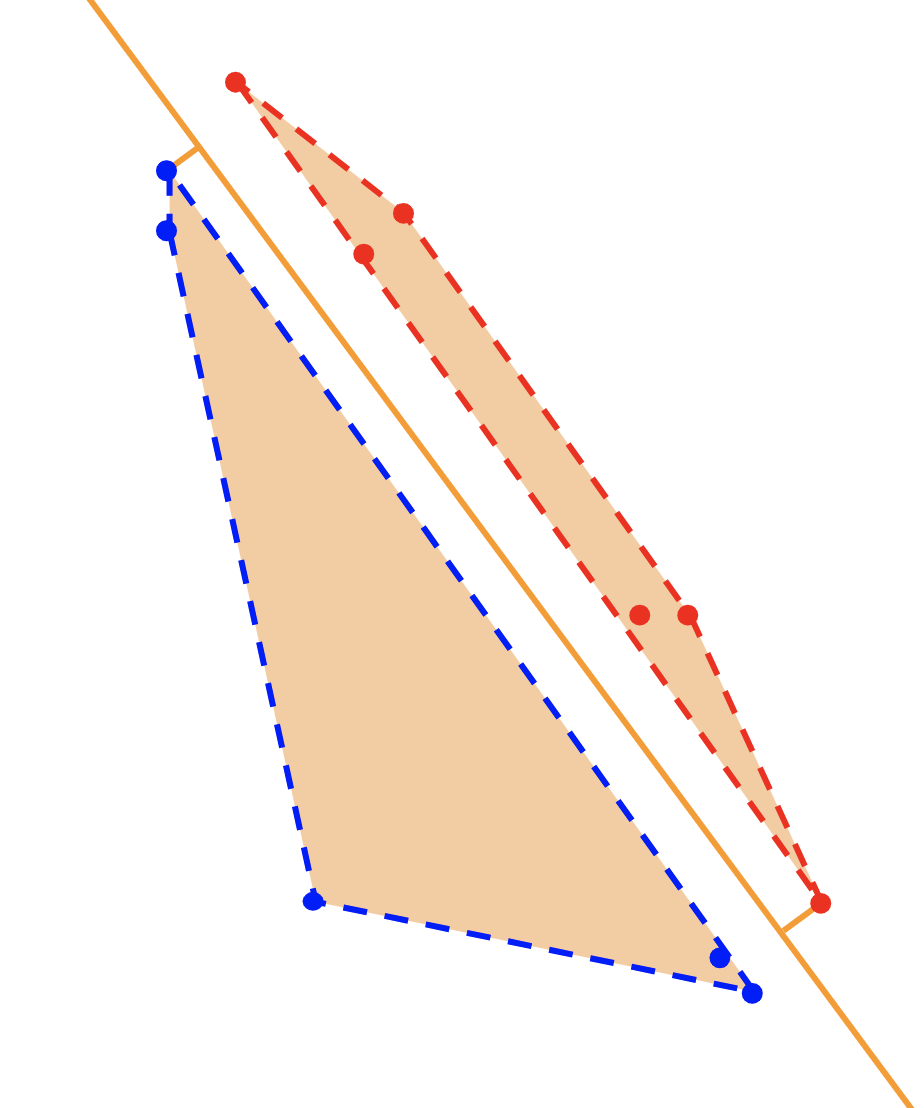}
    \subcaption{}
    \label{fig:separator}
\end{minipage}%
\caption{Congestion between nets denoted by red and blue pins. \textbf{(\subref{fig:rudy})} Rectilinear density metrics are pessimistic. \textbf{(\subref{fig:separator})} An optimistic model of routability. The margin between the convex hulls of nets.}
\label{fig:ec}
\end{figure}

As mentioned in Sec. \ref{sec:related}, typical approaches to model routing congestion rely on estimating or expressing the routed wire density as a function of pin-density and the feasible routing-area (e.g. the pin-bounding box \cite{spindler2007rudy}). The goal is then to minimize this notion of wire-density. In this work, we model the feasible routing region as the convex-hull of the net-pins, and our goal is to separate routing regions. This prevents over-estimation of the routing density as shown in Fig. \ref{fig:ec}. The method consists of two steps: 
\begin{enumerate}
\itemsep=0em
    \item Given two nets, find the max-margin separator $h$.
    \item For all movable components, find directions (gradients) that maximize the margin with respect to $h$. 
\end{enumerate}
Consider a $k$-pin net $e$ defined by pins with coordinates $e_1 = [{e_1}_x, {e_1}_y], e_2 = [{e_2}_x, {e_2}_y], \ldots, e_k = [{e_k}_x, {e_k}_y]\in \mathbb{R}^2$. Note that the coordinates of \textit{any} point lying in the convex hull of the net-area may be written as a convex combination of these pin coordinates. Let $A_e^\top=
\begin{bmatrix}
{e_1}_x & {e_2}_x & \hdots & {e_k}_x  \\
{e_1}_y & {e_2}_y & \hdots & {e_k}_y
\end{bmatrix}$ 
%
be the \textit{pin-matrix} associated with net $e$. Now consider a $k'$-pin net  $e'$ with associated pin matrix $A_{e'}$. 
For the convex hulls characterized by the pins to \textit{not} intersect, there necessarily must be a $u \not = 0$ and $\gamma$ such that $x^\top u - \gamma$ is nonpositive for all $x$ lying in one net, and nonnegative for all $x$ lying in the other net. We denote the hyperplane defined by $\{x | x \in \mathbb{R}^2, x^\top u = \gamma\}$ the \textit{separating hyperplane}, and we want to introduce a regularizer which encourages $e$ and $e'$ to lie in different half-spaces with sufficient margin.
Note that $e$ and $e'$ do not intersect if there is no shared point in the interior of the convex hulls characterized by the coordinates of pins in $e$ and $e'$\textemdash i.e. if the following has \textit{no solution}.
\begin{flalign*}
&\exists \delta_{A_e} \in \mathbb{R}^k, \delta_{A_{e'}} \in \mathbb{R}^{k'} \\ 
&\textrm{ such that } A_e^\top\delta_{A_e} = A_{e'}^\top\delta_{A_{e'}} \quad \mathbf{1}^\top\delta_{A_e}, \mathbf{1}^\top\delta_{A_{e'}}=1,\\ 
&\phantom{\textrm{ such that }}\delta_{A_e}, \delta_{A_{e'}} \geq 0
\end{flalign*}
Namely, duality \& Farkas' Lemma 
provide the conditions that must be satisfied if the convex hulls defined by the pins \textit{do not} intersect:
\begin{flalign*}
&A_{e}u \geq \alpha \mathbf{1} \quad A_{e'}u\leq \beta \mathbf{1} \quad \alpha - \beta > 0 \\
&\implies A_{e}u - \gamma \mathbf{1} \geq \mathbf{1}, \quad A_{e'}u - \gamma \mathbf{1} \leq -\mathbf{1}
\end{flalign*}
This formulation naturally implies the solution to the following minimization problem. Note that one might alternatively aim to find the \textit{maximum-margin} separator. Due to the equivalence with the SVM optimization problem~\cite{cortes1995support,kristindualgeomsvm}, efficient solvers may be employed to recover $\gamma$ and $u$.
\begin{equation}
\begin{aligned}
0 &= \min_{u, \gamma}f(e,e',u,\gamma) \\
&= \min_{u,\gamma}||(-A_{e}u + (\gamma+1)\mathbf{1})_+||_2 + \\ 
&\:\:\phantom{=\min}||(A_{e'}u - (\gamma - 1)\mathbf{1})_+||_2 
\end{aligned} \label{eq:ns}
\end{equation}
%
\noindent Let $A_e$ be the pin matrix corresponding to net $e$. We define the net-separation regularizer:
\begin{equation}
\begin{aligned}
Ns(\cdot) = \frac{1}{|\mathcal{M}|}\sum_{e'\in \mathcal{E}}&\min_{u, \gamma}f(e,e',u,\gamma) 
\end{aligned} \label{eq:nstotal}
\end{equation}
The gradient of $Ns$ can then be recovered with respect to the $i$-th row (pin coordinates) of $A_e$:
$$
\frac{\partial Ns}{\partial (A_e)_i} = \mathds{1}_{(A_e)_i\cdot u \leq \gamma+1}2\langle u, (\gamma+1)\mathbf{1}-A_eu \rangle 
$$
Note that due to the reliance of the above gradient on $\gamma$, and the reliance of $\gamma$ on pin coordinates $A_e$, we adopt an alternating minimization method described in Sec. \ref{sec:formulation}.B.

%% file: Experiments.tex
\section{NS-Place Placement Flow}
\label{sec::flow}
\begin{figure}[h]
\centering
\includegraphics[width=0.4\textwidth]{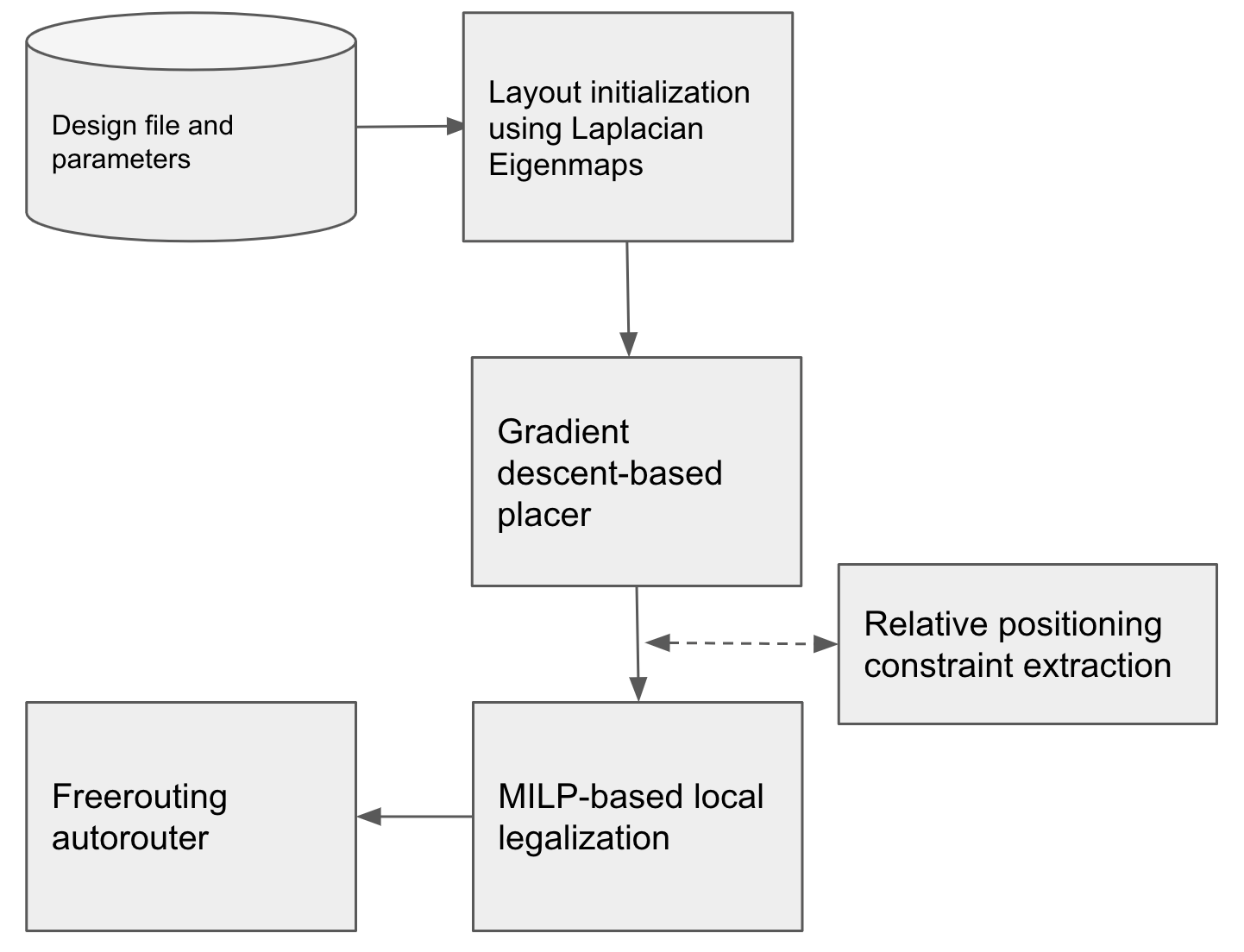}
\caption{Placement procedure with Laplacian Eigenvector initialization, net-separation minimization, and MILP-based legalization with relative positioning constraints.}
\label{fig:flow}
\end{figure}
In this section, we describe the overall flow of our method. The high-level flow is described in Figure \ref{fig:flow}.
\subsection{Initialization with Laplacian eigenvectors}
First-order optimization algorithms are notoriously sensitive to initialization when applied to nonlinear problems. We address this by first collapsing the netlist hypergraph to a component graph via the clique model. We then construct a matrix-representation of the graph connectivity\textemdash the \textit{graph Laplacian}. The solution to the associated eigenvalue problem approximates the solution to the sparsest cut problem \cite{Hall1970}, and clusters arising out of the vertex-projection into the space spanned by the first nontrivial eigenvalues correspond highly connected components of the graph. We use these coordinates to initialize global placement. More concretely, we solve the following problem, where $x$ and $y$ are coordinates of components, $c_i$ are constants, $v$ is a vector of component areas, $V=\mathrm{diag}(v)$, and $L$ is the normalized graph Laplacian; $L = R^{-\frac{1}{2}}AR^{-\frac{1}{2}}$, where $A$ is an adjacency matrix and $R$ is the diagonal degree matrix.
\begin{equation}
\begin{aligned}
    \min\:\: &x^\top Lx + y^\top Ly && &&&  \\
    \textrm{s.t. } &v^\top x = 0,  && v^\top y = 0, &&&  x,y \neq 0 \\
    &x^\top Vx = c_1,  && y^\top Vy = c_2, &&& x^\top Vy = c_3
\end{aligned}
\label{eq:lep}
\end{equation}
Intuitively, the objective is to minimize the weighted squared wire length. The linear and non-equality constraints concentrate the layout about the origin and prevent the trivial solution. The quadratic constraints spread the placement over the $x$ and $y$ axes.

The above problem is unaware of component orientations or fixed position constraints. To resolve this, we generate candidate initializations by considering all possible relative component orientations for a given solution to Problem~\ref{eq:lep}. The solution with minimal cost with respect to Eq.~\ref{eq:obj} is used to seed the global placer.
\subsection{Global placement using coordinate descent}
In Alg.~\ref{alg:nsplace}, we present the detailed steps of our iterative method to reduce congestion.
\begin{algorithm}
\caption{Net crossing minimization}
\begin{flushleft}
\textbf{Input:} Initial placement $[x:y]$, pin matrices $A_e$, $A_{e'}$, regularization parameter $\lambda$ , learning rate $\alpha$, budget $n$ \\
\textbf{Output:} Placement $[x:y]$
\end{flushleft}
\begin{algorithmic}[1]
\Function{NsOpt }{$[x:y],A_e,A_{e'}; \lambda, \alpha$}
\While{Eq.~\ref{eq:obj} not converged}
\State Fix $A_e$, $A_{e'}$, compute $u$, $\gamma$ by Eq. \ref{eq:ns} \Comment{separator}
\State Fix $u$, $\gamma$, compute $\nabla_{p_i}F$, $P$
\State $g \gets \lambda_{D}\nabla_{[x:y]}D + P$ \Comment{compute component update}
\State $[x:y]_{t+1} \gets [x:y]_t + \alpha\cdot g$ \Comment{update components}
\State update $A_e, A_{e'}$ according to $[x:y]_{t+1}$
\EndWhile
\State \Return $[x:y]_{n}$
\EndFunction
\end{algorithmic}
\label{alg:nsplace}
\end{algorithm}

\noindent Recall the proposed global placement optimization problem described in Eq.~\ref{eq:obj}. For brevity, we refer to the cumulative objective as $F$. Note that the above problem may be solved exactly via quadratic programming. We propose to solve the problem approximately by applying first-order methods in an alternating minimization framework by iteratively solving for $u$ and $\gamma$ while keeping $x$ and $y$ fixed and visa-versa. Given a layout, we first compute the separating hyperplane characterized by $u$ and $\gamma$ in Eq.~\ref{eq:ns} (line 4 of Alg.~\ref{alg:nsplace}). Given $u$ and $\gamma$, we can derive the net separation and wirelength gradients associated with individual pins (line 4). To recover the component position update, we introduce the auxiliary variable $P\in \mathbb{R}^{n \times 2}$ corresponding to component update derived from pin-gradients where the $i$-th row of $P$ is defined to be the average of the $i$th component's pin gradients\textemdash i.e. $P_i = \frac{1}{|\mathcal{P}(i)|}\sum_{p \in \mathcal{P}(i)}\nabla_{p}F$. The gradients associated with the density and wirelength terms can then be computed, and component positions updated (lines 5-6). Given these new positions, the pin matrices $A_e$ and $A_{e'}$ can then be updated (line 7).
\subsection{Legalization via mixed integer linear programming}
In this section, we introduce a standard MILP-based layout formulation for placing rectilinear components subject to overlap and boundary constraints. We note that this formulation shares similarities with previous work, e.g. \cite{Funke2016} who discuss the optimality of MILP for wirelength-minimal block placements. Additionally, we integrate relative positioning constraints derived from the global placement procedure to preserve net separation while improving scalability.

\noindent \textbf{MILP-based wirelength-minimal layout}
The objective and constraints are described in the following set of equations. We introduce a second term corresponding to wirelength variance, which we find improves the routability of designs further.
\begin{align*}
    &\min_{x,y,r}\left[\sum_{i\in|\mathcal{E}|}\textrm{hpwl}(i) + \left[\max_{i\in|\mathcal{E}|}\textrm{hpwl}(i) - \min_{i\in|\mathcal{E}|}\textrm{hpwl}(i)\right]\right] \\
    &\:\textrm{hpwl}(i) = (U^{(i)}_x - L^{(i)}_x) + (U^{(i)}_y - L^{(i)}_y)
\end{align*}
where the hpwl term (given for the $x$-direction only) is
\begin{align*}
& U^{(i)}_x \geq p^{(i)}_j(x), \quad L^{(i)}_x \leq p^{(i)}_j(x) \quad \forall j \in \mathcal{E}_i 
\end{align*}
and the solution is subject to non-overlapping constraints (for brevity, boundary constraints are not included):
\begin{align*}
&x_i + r_ih_i+(1-r_i)w_i \leq x_j + W(p_{ij}+q_{ij}) &&\textrm{i-left-j}\\
&y_i + r_iw_i + (1-r_i)h_i \leq y_j + H(1+p_{ij}-q_{ij}) &&\textrm{i-under-j}\\
&x_i - r_jh_j - (1-r_j)w_j \geq x_j - W(1-p_{ij}+q_{ij}) &&\textrm{i-right-j}\\
&y_i - r_jw_j - (1-r_j)h_h \geq y_j - H(2-p_{ij}-q_{ij}) &&\textrm{i-over-j}\\
&x_i,y_i \geq 0, \quad r_i, q_{ij}, p_{ij} \in \{0,1\} && \textrm{variables}
\end{align*}
Where $r_i$, $q_{ij}$, $p_{ij}$ are variables representing orientation and relative positions between modules $i$ and $j$.

\noindent \textbf{Relative positioning constraints}
Given a global placement solution, we derive relative position constraints for pairs of components. By doing so, we preserve the global structure of the global placement solution while allowing the MILP placer to make local adjustments.

For each pair of modules, we consider the minimum horizontal and vertical distances between modules (i.e. from boundary to boundary). We then check if the maximum of the horizontal and vertical distances exceeds a pre-defined threshold $k$ which trades off a preference for HPWL-minimal solutions or routability and runtime. A horizontal or vertical relative position constraint is introduced depending on which distance is greater and the associated binary decision variables ($q_{ij}$ or $p_{ij}$) are removed and the corresponding overlap constraints are simplified or pruned entirely. By integrating these constraints, we preserve the global structure of the analytical solution while eliminating a nontrivial number of decision variables.
\section{Experiments}
\label{sec:results}
%
\input{tables/characteristics.tex}
\input{tables/preroute_results}
\input{tables/postroute_results}
%
\subsection{Experiment setup}
We applied our method to 14 PCB benchmarks~\cite{lincomplicated}. Details are provided in Table~\ref{ref::characteristics}. We solve Eq.~\ref{eq:nstotal} via gradient descent with momentum with $\alpha=1e^{-3}$, $\lambda_{\textrm{NS}}=1$, and $\lambda_{\textrm{D}}=1$. GDMILP corresponds to running the NS-Place with $\lambda_{\textrm{NS}} = 0$, and MILP corresponds to solving the MILP without relative-position constraints. Layout quality is evaluated using the open-source FreeRouting router~\cite{freerouting}. The routed wirelength, number of design rule violations, and the number of vias are reported. Experiments are performed with an 3.4GHz Intel i7-4770 CPU and 31GB RAM. If no optimal solution is found after 4 hours, we report results for the best-so-far (with respect to the MILP objective) feasible solution. If no feasible solution is found, associated entries are marked with "-".
\subsection{Main experiments}
%
%
%
\begin{figure}[ht]
\centering
\begin{subfigure}[b]{0.498\linewidth}
\resizebox{4cm}{4cm}{\scalebox{1.0}[1.0]{\includegraphics[width=\linewidth]{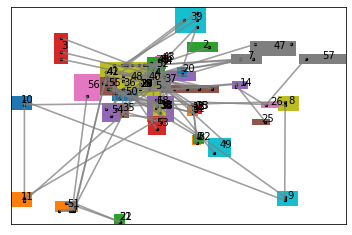}}}
\subcaption{}
\label{fig:lapl}
\end{subfigure}
\begin{subfigure}[b]{0.48\linewidth}
\resizebox{4cm}{4cm}{\scalebox{1.0}[1.062]{\includegraphics[width=\linewidth]{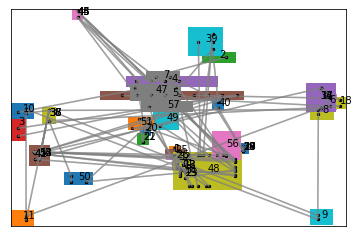}}}
\subcaption{}
\label{fig:graddesc}
\end{subfigure}
\begin{subfigure}[b]{0.498\linewidth}
\resizebox{4cm}{4cm}{\scalebox{1.0}[1.0]{\includegraphics[width=\linewidth]{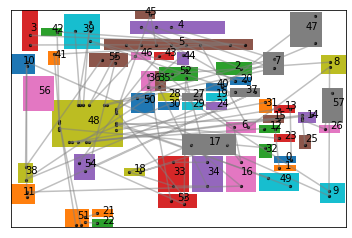}}}
\subcaption{}
\label{fig:milp}
\end{subfigure}
\begin{subfigure}[b]{0.48\linewidth}
\resizebox{4cm}{4cm}{\scalebox{1.0}[1.045]{\includegraphics[width=\linewidth]{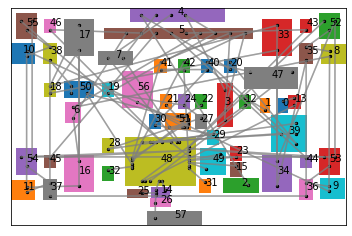}}}
\subcaption{}
\label{fig:manual}
\end{subfigure}
\caption{\textbf{PCB12 layouts}. (\textbf{\subref{fig:lapl}}): Seed placement produced from Laplacian eigenvectors. (\textbf{\subref{fig:graddesc}}) Global placement to minimize net crossings. (\textbf{\subref{fig:milp}}) MILP-based legalization. (\textbf{\subref{fig:manual}}) Manual layout.}
\label{fig:bbbc}
\end{figure}
%
%
%
\noindent \textbf{PCB placement metric comparison}
We provide manual, MILP, GDMILP, and NS-Place placement results in table \ref{tab:preroute_mark}. We show that NS-Place reduces the net separation cost by 77\% and 41\% on average compared to manual and GDMILP. This implies that although the MILP-based fine-tuning step does not optimize net separation explicitly, satisfaction of relative position constraints results in preservation of the global structure produced by the net separation step. Furthermore, the inclusion of the net separation term does not result in a large increase in HPWL. Although the HPWL of NS-Place solutions increase 9\% on average compared to GDMILP, NS-Place still achieves 20\% lower HPWL compared to manual layouts.

We report the cumulative runtime to complete a placement in Table~\ref{tab:preroute_mark}. Note that methods employing relative positioning constraints (i.e., GDMILP, NS-Place) strictly improve over vanilla MILP. We see that NS-Place closely matches the performance of the GDMILP flow with only around 4\% increase on average runtime. This implies that the net separation regularizer imposes low overhead. Fig.~\ref{fig:bbbc} (a), (b), and (c) show the PCB placement results of PCB12 of each stage of automatic placement flow. Fig.~\ref{fig:bbbc} (d) is the manual layout of PCB12. Compared to Fig.~\ref{fig:bbbc} (d), we observe that NS-Place produces placement with fewer net crossings as shown in Fig.~\ref{fig:bbbc} (c).

\noindent \textbf{PCB routing metric comparison}
\begin{figure}[!ht]
\centering
\begin{subfigure}[b]{0.498\linewidth}
\begin{tikzpicture}
\node[anchor=south west,inner sep=0] at (0,0) {
\reflectbox{\rotatebox[origin=c]{180}{\includegraphics[width=\linewidth,trim=4 25 4 4,clip]{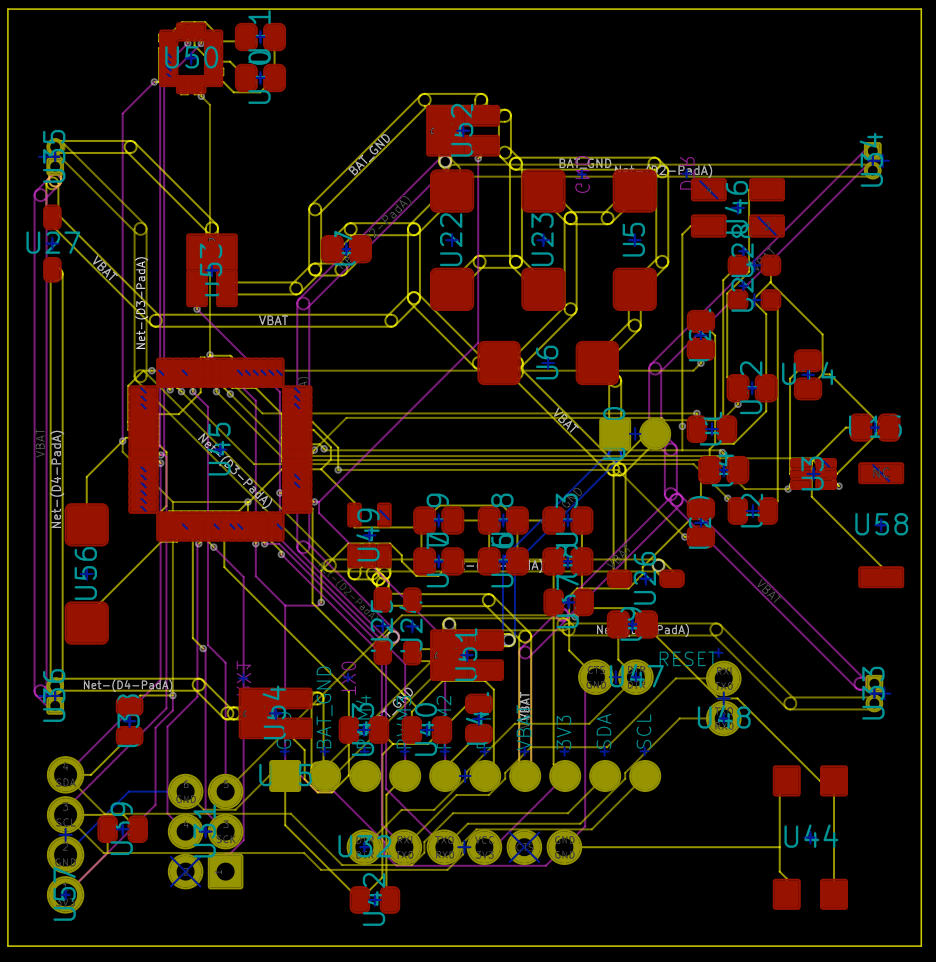}}}};
\draw[green,ultra thick,rounded corners] (1.5,1.8) rectangle (2.8,2.4);
\draw[green,ultra thick,rounded corners] (0.5,2.6) rectangle (1.5,3.2);
\end{tikzpicture}
\caption{}
\label{fig:milprouted}
\end{subfigure}
\begin{subfigure}[b]{0.48\linewidth}
\reflectbox{\rotatebox[origin=c]{180}{\includegraphics[width=\linewidth]{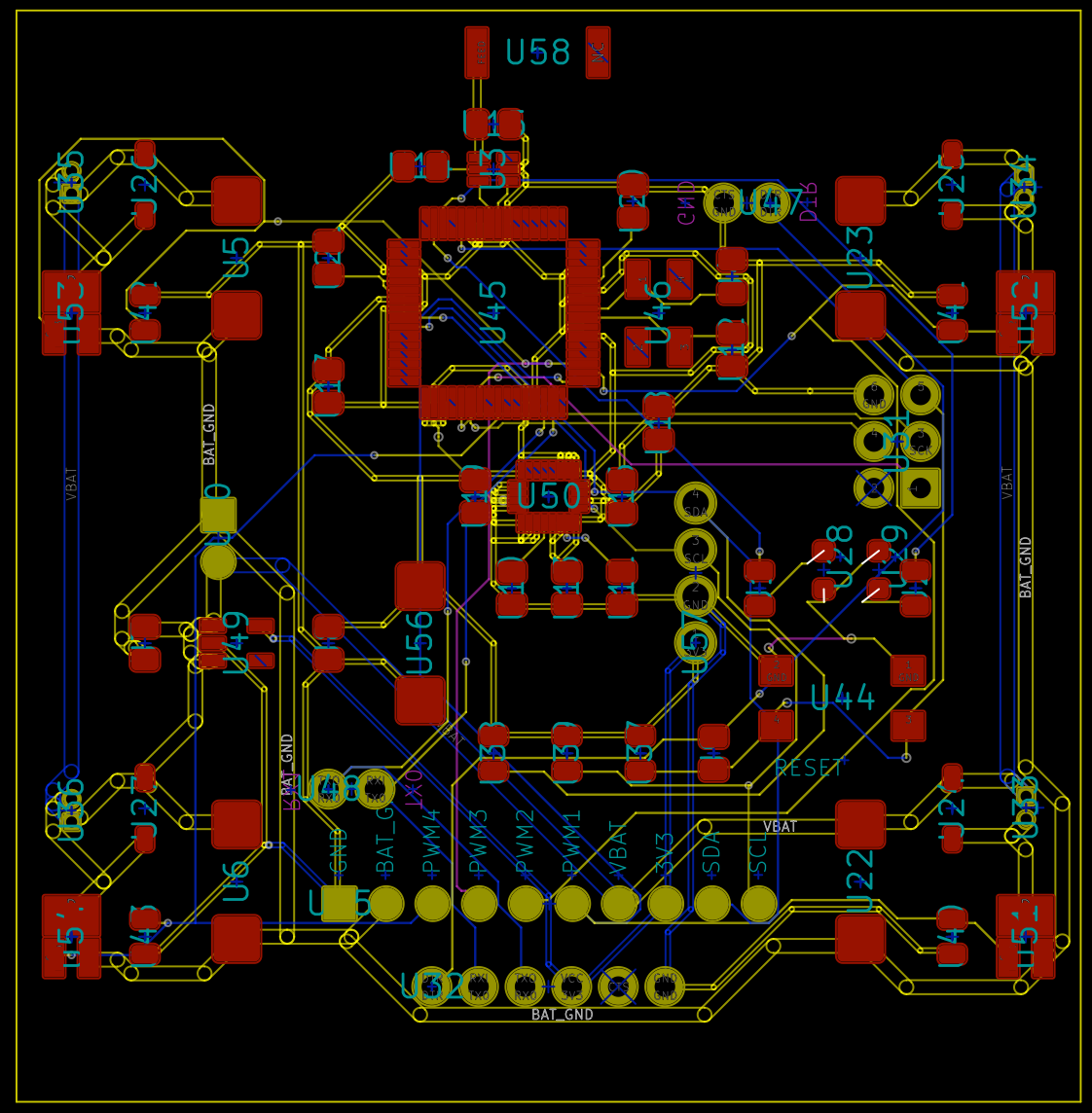}}}
\caption{}
\label{fig:manrouted}
\end{subfigure}
\caption{\textbf{PCB12 P\&R result in KiCAD}. (\textbf{\subref{fig:milprouted}}) Solution produced by our method. Routable regions are emphasized with green rectangles. (\textbf{\subref{fig:manrouted}}) Routed manual placement.}
\label{fig:bbbc1}
\end{figure}
We evaluate routability in Table~\ref{tab:postroute} by reporting the routed wirelength, the number of DRVs and vias, and the number of unrouted nets using FreeRouting and Kicad. Compared with manual, MILP, and GDMILP, NS-Place reduces the \#DRVs and \#unrouted nets by roughly 80\%, 70\%, and 75\% on average.

NS-Place additionally reduces the routed wirelength by 10\% and 6\% on average compared to manual and GDMILP respectively and reduces the \#Vias by 18\% and 39\%. Moreover, for PCBs 7, 10, 11, and 13, which have \#components larger than 60 and \#pins larger than 300, NS-Place achieves 34\% \#Vias reduction on average compared to manual. An example routed result is given in Fig.~\ref{fig:bbbc1}. Compared to the manual layout, NS-Place successfully improves routability via net separation (i.e., the routable regions near the center of the board). Note that NS-Place fails to produce a superior solution for PCB2. We attribute this to PCB2's high utilization, with only about 20\% of the board area available to the global placer as whitespace. In summary, we demonstrate that NS-Place significantly reduces \#DRVs, \#unrouted nets, and \#Vias for PCB reliability compared to manual, MILP, and GDMILP with an extensive study on 14 PCBs.

%% file: tables/characteristics.tex
\begin{table}[htb]
\small
\centering
\caption{\label{tab:designs}Design characteristics. locked are fixed components. layers are layers available for routing.}
\resizebox{0.48\textwidth}{!}{%
\begin{tabular}{@{}rrrrrrrr@{~~}}
\toprule
design &  $W\times H (mm^2)$ &  \#comp & \#locked. & util.  & \#nets & \#pins & \#layers\\
\hline
\rowcolor[HTML]{EFEFEF}
PCB1    &    $21\times 14$      &  8       & 1   &   0.59      &   15 & 40  &  1   \\
PCB2    &    $51\times 23$     &  18       &  5  &   0.79      &   34 &  77 &  2 \\
\rowcolor[HTML]{EFEFEF}
PCB3    &    $55\times 28$     &  34       & 2   &   0.44       &    38 &  138 & 2  \\
PCB4    &    $23\times 60$      &  28       & 6    &   0.67       &   52  & 140 &  2  \\
\rowcolor[HTML]{EFEFEF}

PCB5    &    $41\times 22$      &  48       & 2    &   0.40       &   54   &  163 & 2  \\
PCB6    &    $62\times 57$      &  48       & 2    &   0.17       &   64   & 190  & 2 \\
\rowcolor[HTML]{EFEFEF}
PCB7    &    $51\times 23$      &  46       & 2    &   0.55       &   69  &  211  & 2 \\
PCB8    &    $57\times 87$      &  36       & 2    &   0.62       &   70   &  188 & 2 \\
\rowcolor[HTML]{EFEFEF}
PCB9    &   $44\times 36$       &  58       & 2    &   0.60       &   80   &  229 & 2 \\
PCB10    &    $102\times 54$      &  57       & 18    &   0.21       &   99 & 319 & 2   \\
\rowcolor[HTML]{EFEFEF}
PCB11    &    $89\times 58$      &  64       & 2    &   0.10       &   134   &  401 & 4  \\
PCB12    &    $58\times 60$      &  58       & 4    &   0.31       &   35   &  233 & 4 \\
\rowcolor[HTML]{EFEFEF}
PCB13    &    $86\times 72$      &  61       & 4    &   0.51       &   63  &  314 & 4 \\
PCB14    &    $86\times 54$      &  1570     & 947    &   0.64       &   386  &  1638 & 4 \\

\bottomrule
\end{tabular}\par
}
\label{ref::characteristics}
\end{table}

%% file: tables/preroute_results.tex
\begin{table*}[ht]
\caption{Pre-route metrics for PCB designs. We report the cumulative HPWL and the net separation cost. The top performing result is \textbf{bolded}. The "-" represents that MILP cannot produce a feasible placement in 4 hours.}
\label{tab:preroute_mark}
\tabcolsep = 4pt
\centering
\resizebox{0.98\textwidth}{!}{%
\begin{tabular}{|c|r|r|r|r|r|r|r|r|r|r|r|}
\hline
\multicolumn{1}{|l|}{\multirow{2}{*}{Design}} & \multicolumn{4}{c|}{HPWL (mm)}                                                                                        & \multicolumn{4}{c|}{Net Separation Obj.}                                                                              & \multicolumn{3}{c|}{Runtime (s)}                                                        \\ \cline{2-12} 
\multicolumn{1}{|l|}{}                        & \multicolumn{1}{c|}{manual} & \multicolumn{1}{c|}{MILP} & \multicolumn{1}{c|}{GDMILP} & \multicolumn{1}{c|}{NS-Place} & \multicolumn{1}{c|}{manual} & \multicolumn{1}{c|}{MILP} & \multicolumn{1}{c|}{GDMILP} & \multicolumn{1}{c|}{NS-Place} & \multicolumn{1}{c|}{MILP} & \multicolumn{1}{c|}{GDMILP} & \multicolumn{1}{c|}{NS-Place} \\ \hline
PCB1                                          & 110.22                      & \textbf{64.44 (41.50)}             & 68.90 (37.4)                & 72.10 (34.60)                 & 192.13                      & 192.11                    & 192.11                      & \textbf{187.40}                        & 2.70                      & 4.90                        & 5.10                          \\ \hline
PCB2                                          & 362.70                      & \textbf{291.10  (19.70)}           & 294.32 (18.80)              & 296.18 (18.30)                & 497.50                      & 497.40                    & 499.30                      & \textbf{493.10}                        & 6.30                      & 6.20                        & 6.80                          \\ \hline
PCB3                                          & 312.60                      & \textbf{212.70 (32.0)}             & 252.40 (19.30)              & 271.30 (13.20)                & 1427.10                     & 2124.30                   & 1426.40                     & \textbf{1411.40}                       & 14400.00                  & 3793.00                     & 3917.60                       \\ \hline
PCB4                                          & 603.30                      & -                         & 536.48 (11.10)              & \textbf{531.19 (12.00)}                & 3750.20                     & -                         & 3753.10                     & \textbf{3746.13}                       & -                         & 9.40                        & 9.90                          \\ \hline
PCB5                                          & 654.10                      & -                         & \textbf{621.00 (5.10)}               & 637.00 (2.60)                 & 2400.00                     & -                         & 2400.00                     & \textbf{2300.00}                       & -                         & 4231.20                     & 4461.50                       \\ \hline
PCB6                                         & 771.80                      & -                         & \textbf{649.40 (15.90)}              & 708.60 (8.40)                 & 1930.00                     & -                         & 1940.00                     & \textbf{1740.00}                       & -                         & 4893.10                     & 5072.40                       \\ \hline
PCB7                                          & 2987.90                     & -                         & \textbf{563.10 (88.50)}             & 1601.40 (46.40)               & 1241.34                     & -                         & 1239.70                     & \textbf{1017.16}                       & -                         & 5927.30                     & 6008.20                       \\ \hline
PCB8                                         & 766.10                      & \textbf{731.30 (4.60)}             & 748.40 (2.30)               & 756.40 (1.30)                 & 98500.00                    & 99500.00                  & 99400.00                    & \textbf{96400.00}                      & 14400.00                  & 4360.90                     & 4732.40                       \\ \hline
PCB9                                          & 714.90                      & -                         & 677.60 (5.20)               & \textbf{662.50 (7.30)}                 & 2600.00                     & -                         & 2600.00                     & \textbf{2400.00}                       & -                         & 5327.30                     & 5719.80                       \\ \hline
PCB10                                          & 4355.61                     & -                         & 3317.94 (23.80)             & \textbf{3315.46 (23.90)}               & 2117.80                     & -                         & 2124.10                     & \textbf{2103.60}                       & -                         & 5314.70                     & 5417.20                       \\ \hline
PCB11                                         & 2941.90                     & -                         & \textbf{2573.20 (12.50)}             & 2618.10 (11.00)               & 9210.00                     & -                         & 9210.00                     & \textbf{9070.00}                       & -                         & 4651.90                     & 4782.40                       \\ \hline
PCB12                                          & 972.50                      & 929.30 (4.40)             & \textbf{932.60 (4.10)}               & 941.30 (3.20)                 & 10.73                       & 11.31                     & 11.47                       & \textbf{36.92}                         & 14400.00                         & 4619.10                     & 4752.30                       \\ \hline
PCB13                                         & 2644.97                     & -                         & \textbf{2126.13 (19.60)}             & 2151.77 (18.60)               & 2400.00                     & -                         & 2400.00                     & \textbf{2300.00}                       & -                         & 3278.60                     & 3416.30                       \\ \hline
PCB14                                       &         7069.26               & -                         &     \textbf{6432.19 (9.01)}        &      6691.34 (5.35)         &        29400            & -                         &      29870                &          \textbf{11186}             & -                         &           14400.00           &         14400.00               \\ \hline 

\end{tabular}
}
\end{table*}

%% file: tables/postroute_results.tex
\begin{table*}[ht]
\caption{Post-route metrics for PCB designs. We report routed wirelength using FreeRouting, the number of vias, and the number of DRVs as reported by KiCAD. We report the percent improvement in parenthesis. The best result is \textbf{bolded}.}
\label{tab:postroute}
\tabcolsep = 4pt
\centering
\resizebox{0.98\textwidth}{!}{%
\begin{tabular}{|c|r|r|r|r|r|r|r|r|r|r|r|r|}
\hline
\multicolumn{1}{|l|}{\multirow{2}{*}{Design}} & \multicolumn{4}{c|}{Routed Wirelength (mm)}                                                                           & \multicolumn{4}{c|}{\#Vias}                                                                                           & \multicolumn{4}{c|}{\#DRVs + \#unrouted nets}                                                                        \\ \cline{2-13} 
\multicolumn{1}{|l|}{}                        & \multicolumn{1}{c|}{manual} & \multicolumn{1}{c|}{MILP} & \multicolumn{1}{c|}{GDMILP} & \multicolumn{1}{c|}{NS-Place} & \multicolumn{1}{c|}{manual} & \multicolumn{1}{c|}{MILP} & \multicolumn{1}{c|}{GDMILP} & \multicolumn{1}{c|}{NS-Place} & manual                     & \multicolumn{1}{c|}{MILP} & \multicolumn{1}{c|}{GDMILP} & \multicolumn{1}{c|}{NS-Place} \\ \hline
PCB1                                          & 129.00                      & 123.00 (4.70)             & 194.00 (-50.4)              & \textbf{121.00 (6.20)}                 & \textbf{0}                           & 10                        & 9                           & 4                             & 6                          & \textbf{0}                         & \textbf{0}                           & \textbf{0}                             \\ \hline
PCB2                                          & \textbf{354.00}                      & 632 (-78.50)              & 507.00 (-43.20)             & 421.00 (-18.90)               & \textbf{3}                           & 6                         & 8                           & 5                             & 13                         & 26                        & 24                          & 23                            \\ \hline
PCB3                                          & 638.00                      & 682.00 (-6.90)            & 771.00 (-20.80)             & \textbf{616.00 (3.40)}                 & 17                          & 21                        & 24                          & \textbf{15}                            & 11                         & 2                         & 3                           & \textbf{0}                             \\ \hline
PCB4                                          & 809.00                      & -                         & 857.00 (-5.90)              & \textbf{806.00 (0.40)}                 & \textbf{31}                          & -                         & 59                          & 54                            & 17                         & -                         & 4                           & \textbf{0}                             \\ \hline
PCB5                                          & 558.00                      & -                         & \textbf{538.00 (3.60)}               & 541.00 (3.00)                 & \textbf{32}                          & -                         & 84                          & 49                            & 7                          & -                         & 5                           & \textbf{3}                             \\ \hline
PCB6                                         & 1007.00                     & -                         & \textbf{854.00 (15.90)}              & 906.00 (10.00)                & 23                          & -                         & 47                          & \textbf{19}                            & 34                         & -                         & 29                          & \textbf{0}                             \\ \hline
PCB7                                          & 3735.00                     & -                         & 3140.00 (15.90)             & \textbf{2949.00 (21.00)}               & 161                         & -                         & 141                         & \textbf{93}                            & 23                         & -                         & \textbf{0}                           & \textbf{0}                             \\ \hline
PCB8                                         & 913.00                      & 854.30 (6.30)             & 713.40 (21.90)              & \textbf{711.90 (22.00)}                & 57                          & 73                        & 89                          & \textbf{43}                            & 27                         & 6                         & 5                           & \textbf{0}                             \\ \hline
PCB9                                          & 1069.00                     & -                         & \textbf{749.00 (29.90)}              & 772.00 (27.80)                & \textbf{38}                          & -                         & 87                          & 64                            & 17                         & -                         & 7                           & \textbf{0}                             \\ \hline
PCB10                                          & 5043.00                     & -                         & 5294.00 (-1.00)             & \textbf{4914.00 (2.60)}                & 129                         & -                         & 136                         & \textbf{97}                            & 12                         & -                         & 3                           & \textbf{0}                             \\ \hline
PCB11                                         & 3460.00                     & -                         & 3271.90 (5.40)              & \textbf{3107.80 (10.20)}               & 131                         & -                         & 286                         & \textbf{109}                           & 23                         & -                         & 57                          & \textbf{13}                            \\ \hline
PCB12                                          & 1790.00                     & 1960.00 (-9.40)           & 1930.00 (7.80)             & \textbf{1720.00 (3.90)}                & 49                          & 62                        & 54                          & \textbf{45}                            & 4                          & 9                         & 8                           & \textbf{0}                             \\ \hline
PCB13                                         & 3150.00                     & -                         & 2903.00 (7.80)              & \textbf{2897.00 (8.00)}                & 161                         & -                         & 98                          & \textbf{83}                            & 11                         & -                         & 9                           & \textbf{0}                             \\ \hline
PCB14                                         &    9017.00                  & -                         &   9643.00 (-6.94)            &    \textbf{8873.00 (1.60)}            &          976                & -                         &            1007              &      \textbf{942}                      &         104                 & -                         &     139                      &        \textbf{92}                     \\ \hline 
\end{tabular}
}
\end{table*}

%% file: Conclusion.tex
\section{Conclusion and Future Work}
\label{sec:conclusion}
We have presented a novel algorithm which encourages routability of multi-pin nets in PCB designs. 
In an extensive study on 14 PCB designs, we have demonstrated that NS-Place achieves 80.98\%, 70.00\%, and 74.68\% reduction on average \#DRVs and \#unrouted nets for routability and 34.36\% fewer \#Vias on average. We plan to extend our technical and empirical analysis to integrated circuit (IC) placement and explore congestion minimization via nonlinear separators and plan to investigate methods to optimize component orientation during the net separation minimization procedure. 
Finally, while first-order methods are efficient in practice, due to the nonlinearity of our objective, optimal solutions are not guaranteed and initialization remains an open problem.

\section{Acknowledgements}
\label{sec:acknowledgement}
We acknowledge the support of the the DARPA IDEA Project and National Science Foundation CCF-2110419.